\documentclass{aa}  

\usepackage{graphicx}
 \usepackage{hyperref}
 \usepackage{txfonts}
 \usepackage{natbib}
\usepackage{multirow}
\newcommand{\Rhill}{{R_{\rm Hill}}}
\newcommand{\ainn}{{a_{\rm inn}}}
\newcommand{\aout}{{a_{\rm out}}}
\newcommand{\asin}{{a_{\rm sin}}}
\newcommand{\einn}{{e_{\rm inn}}}
\newcommand{\eout}{{e_{\rm out}}}
\newcommand{\esin}{{e_{\rm sin}}}
\newcommand{\iinn}{{i_{\rm inn}}}

\newcommand{\isin}{{i_{\rm sin}}}
\newcommand{\msmbh}{{M_{\rm SMBH}}}
\newcommand{\mout}{m_{\rm out}}

\newcommand{\sset}[1]{\texttt{$\sigma$-#1}}
\newcommand{\sszero}{\texttt{$\sigma$0}{}}

\newcommand{\msun}{{\mathrm{M}_\odot}}
\newcommand{\au}{{\mathrm{au}}}
\newcommand{\pc}{{\mathrm{pc}}}
\newcommand{\Renc}{{\mathrm{R_{\rm enc}}}}

\newcommand{\REV}[1]{{#1}}
\newcommand{\REVV}[1]{{#1}}
\usepackage[normalem]{ulem}

\begin{document} 

   \title{Three-body encounters in black hole discs\\ around a supermassive black hole:}

   \subtitle{The disc velocity dispersion and the Keplerian tidal field determine the eccentricity and spin-orbit alignment of gravitational wave mergers}

   \author{A.A.Trani
          \inst{1}\fnmsep\inst{2}\fnmsep\inst{3}%\thanks{Just to show the usage of the elements in the author field}
          ,
          S. Quaini\inst{4}
          \and
          M. Colpi\inst{4,5}
          }
   \institute{Niels Bohr International Academy, Niels Bohr Institute, Blegdamsvej 17, 2100 Copenhagen, Denmark
     \and Research Center for the Early Universe, School of Science, The University of Tokyo, Tokyo 113-0033, Japan
     \and Okinawa Institute of Science and Technology, 1919-1 Tancha, Onna-son, Okinawa 904-0495, Japan   
    \email{aatrani@gmail.com}
     \and
      Dipartimento di Fisica 'G. Occhialini', Università degli Studi di Milano-Bicocca, Piazza della Scienza 3, I-20126 Milano, Italy
      \and
      INFN, Sezione di Milano-Bicocca, Piazza della Scienza 3, 20126 Milano, Italy\\
             }
   \date{Received XX XX, XXXX; accepted XX XX, XXXX}

% \abstract{}{}{}{}{} 
% 5 {} token are mandatory
 
  \abstract
   {Dynamical encounters of stellar-mass black holes (BHs) in a disc of compact objects around a supermassive BH (SMBH) can accelerate the formation and  coalescence of BH binaries. It has been proposed that binary--single encounters among BHs in such discs can lead to an excess of highly eccentric BH mergers. However, previous studies have neglected how the disc velocity dispersion and the SMBH's tidal field affect the three-body dynamics.}
   {We investigate the outcomes of binary--single encounters considering  different values of the disc velocity dispersion, and examine the role of the SMBH's tidal field. We then demonstrate how their inclusion affects the properties of merging BH binaries.}
   {We performed  simulations of four-body encounters (i.e. with the SMBH as the fourth particle) using the highly accurate, regularised code \textsc{tsunami}, which includes post-Newtonian corrections up to order 3.5PN. To isolate the effect of the SMBH's tidal field, we compared these simulations with those of three-body encounters in isolation.}
   {
   The disc velocity dispersion controls how orbits in the disc are aligned and circular, and determines the relative velocity of the binary--single pair before the encounter. As the velocity dispersion decreases, the eccentricity of post-encounter binaries transitions from thermal to superthermal, and binaries experience enhanced hardening. The transition between these two regimes happens at disc eccentricities and inclinations of order $e \sim i \sim 10^{-4}$. \REVV{These distinct regimes correspond to a disc dominated by random motions ($e \sim i \gtrsim 10^{-4}$) and one dominated by the Keplerian shear ($e \sim i \lesssim 10^{-4}$)}.
   The effect of the SMBH's tidal field depends on the velocity dispersion of the disc.
   When the velocity dispersion is low, the resulting binaries are less eccentric compared to isolated encounters. Conversely, binaries become less eccentric compared to isolated encounters at high velocity dispersion. This also affects the number of BH mergers. 
  }
   {The inclusion of the SMBH's tidal field and the disc velocity dispersion can significantly affect the number of GW mergers, and especially the number of highly eccentric inspirals.  These can be up to ${\sim}2$ times higher at low velocity dispersion, and ${\sim}12$ times lower at high velocity dispersions. The spin--orbit alignment is influenced by the tidal field exclusively at high velocity dispersions, effectively inhibiting the formation of anti-aligned binary BHs.
   \REVV{Nonetheless, encounters in random-motion-dominated discs around a SMBH  are still more effective in producing GW mergers compared to those occurring in spherically symmetric nuclear star clusters without an SMBH}.
}


   \maketitle
%
%-------------------------------------------------------------------

\section{Introduction}
Discs of stellar-mass black holes (BHs) around supermassive BHs (SMBHs)  may be common in the Universe, as suggested by both observations (albeit indirectly) and theoretical analysis.
On the observational side, Wolf-Rayet and O/B stars have been observed orbiting SgrA* in one or more discs \citep{gen03a,lev03,pau06,lu09,bar09,yel14,vonfellenberg2022}, which, at the end of their stellar lifecycle, will leave a disc of compact objects. Additionally, a disc of early-type stars has been observed around the SMBH in M31, which suggests that ubiquitous mechanisms may form nuclear discs  \citep{lauer1993,tremaine1995,peiris2003,lauer2005,bender2005,brown2013}.

Numerical studies have proposed a variety of ways to form nuclear discs around SMBHs, such as fragmentation of tidally disrupted molecular clouds \citep{map12,tra16,trani2018,mastrobuono-battisti2019}, gas funnelling during galaxy mergers \citep{hopkins2010a,hopkins2010b}, anisotropic mass segregation \citep{slogien2019,kamlah2022,mathe2023}, globular cluster infall and disruption \citep{arca-sedda2015,arca-sedda2018}, star formation in active galactic nuclei (AGN) discs \citep{cantiello2021,jermyn2022}, or gas-friction capture in AGN discs \citep{yang2019,fabj2020,mcleod2020,nasim2022,generozov2023}. In the last two cases, the nuclear disc is embedded into the AGN's gaseous disc, which is in contrast to dry nuclear discs formed by the other processes. 

Discs of compact objects around an SMBH are excellent candidate environments for producing gravitational wave (GW) sources for both ground-based and space-borne GW detectors \citep{gwtc-1,gwtc2020a,gwtc2020b,gwtc-2.1,gwtc-3,gwtc-3pop}. In particular, the high stellar density in galactic nuclei makes them rich in dynamical interactions, which can give rise to unique signatures in the GW waveform that can be used to infer their origin \citep[see e.g.][]{mapelli2021review,speratrani2022}.
For instance, the gravitational interaction of BH binaries with the SMBH may give rise to oscillations in the binary eccentricity, called von~Zeipel-Kozai-Lidov (ZKL) oscillations, which can accelerate and trigger GW coalescence of the binary \citep{zeipel1910,lid62,koz62,blaes2002,miller2009,antonini2012,liu2015,naoz2016,shev2017,hoa18,martinez2021,trani2022}. 
In addition to ZKL oscillations, gravitational interactions might occur in the disc itself in the form of few-body encounters amongst single objects and binaries. Such interactions can lead to single--single GW captures or accelerate the formation and merger of BH binaries \citep{bartos2017,leigh2018,secunda2019,secunda2021,lijiaru2022,boekholt2023,delaurentis2023,rowan2023,lijiaru2023,rom2024,qiankecheng2024}. \citet{samsing2022} argued that binary--single encounters in AGN discs may give rise to highly eccentric mergers because of the low relative velocity between bodies in the disc. However, their study did not consider the effect of the tidal field of the  SMBH on the three-body dynamics in the disc. The role of this tidal field was first studied by \citet{trani2019b} in the context of binary encounters with BHs from an isotropic cusp, but a detailed study on how it affects binary--single encounters in a disc is still missing.

In this paper, we focus on the outcome of three-body encounters between single and binary BHs in a dry nuclear disc around an SMBH by means of highly accurate few-body simulations.
Particularly, we describe (i) the effect of different disc velocity dispersions on the binary outcomes (Section~\ref{sec:vdisc}) and (ii) the effect of the SMBH tidal field compared to three-body encounters in its absence (Section~\ref{sec:smbhtid}). We put our findings into the context of GW sources in Section~\ref{sec:gw}. Our numerical methods and simulation setup are described in Section~\ref{sec:num}.

%--------------------------------------------------------------------
\section{Numerical setup}\label{sec:num}

\subsection{Initial conditions}\label{sec:ics}

\begin{figure}
        \centering
        \includegraphics[width=\columnwidth]{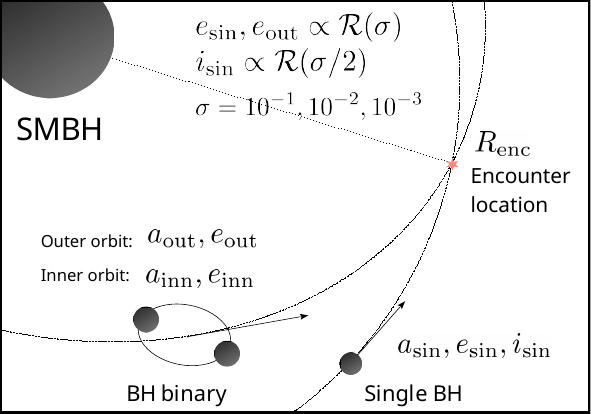}
        \caption{Scheme of the initial setup of our simulations. The centre of mass of the binary and the single are set to encounter at a location $R_{\rm enc}$ with impact parameter $b \in (0, 2 \ainn)$. The eccentricities and mutual inclinations of the two orbits around the SMBH are drawn from a Rayleigh distribution ($\cal R$) with scale parameter $\sigma$ and $\sigma/2$, respectively.
        }
        \label{fig:scheme}%
\end{figure}

We run four-body simulations of an encounter between a BH binary and a single BH around an SMBH, assuming that both objects are part of the same nuclear disc. The initial setup of the system is outlined in Figure~\ref{fig:scheme}.
The binary orbits around the SMBH with a semi-major axis and eccentricity $\aout$ and $\eout$ (hereafter  we refer to this as the outer orbit).
Likewise, the binary is characterised by an `inner'  semi-major axis and inner eccentricity $\ainn$ and $\einn$, respectively. The orbit of the  single BH around the SMBH has a semi-major axis $\asin$ and eccentricity $\esin$ and is inclined by $\isin$ with respect to the orbital plane of the binary. 

The eccentricities $\eout$ and $\esin$ are drawn from the same distribution, which is consistent with the binary and the single being part of the same disc, while $\isin$ relates to the opening angle of the disc. More generally, eccentricity and inclination express the deviation from an aligned circular orbit, and parametrise the random motion of Keplerian orbits in a disc in the same way as the velocity dispersion describes the random motion of stars in an isotropic stellar cluster. Specifically, the random velocity is related to $e$ and $i$ as \citep{lissauer1993}
\begin{equation}\label{eq:eivdisc}
v \approx (e^2 + i^2)^{1/2} \, v_{\rm K}        \:,
\end{equation}
where $v_{\rm K}$ is the Keplerian circular velocity. \REV{The random motions of particles in a disc will grow because of two-body gravitational interactions, keeping the ratio $\langle e \rangle^{1/2} / \langle i\rangle^{1/2} \approx 2$ approximatively constant, with eccentricities and inclinations growing with time as $e, i \propto t^{1/4}$. In this regime, both the inclination and eccentricity distributions are accurately modelled by a Rayleigh distribution, a finding corroborated by both simulations and theoretical analyses \citep{ida1992,stewart2000,kokubo2002}. Specifically, a Rayleigh distribution in Keplerian orbital elements results from assuming that the random velocity field is described by a triaxial Gaussian distribution. This is only valid in the regime of small inclinations and eccentricities, where the orbits are well approximated by linear epicycles.} 

For this reason, we parametrise the disc velocity dispersion with a single dimensionless parameter $\sigma \equiv \sigma_e = 2 \sigma_i$, where $\sigma_e$ and $\sigma_i$ are the scale parameters of the Rayleigh distributions ($\cal R$) of eccentricity and inclination. We choose $\sigma = 10^{-1}, 10^{-2}, 10^{-3}, 10^{-4}, 10^{-5}, 10^{-6}, 10^{-7}, 0$,  corresponding to simulation sets $\sigma n$, where $n$ goes from $-1$ to $-7$, and then to $0$. The sets correspond to eccentricity and inclinations of order $(e, i) \sim (1.25, 36^\circ) \times \sigma$, which is going from $e,i \sim 0.125, 3.6^\circ$ for \sset{1}, to $e,i \sim 1.25\times 10^{-7}, 3.6^\circ\times 10^{-7}$ for \sset{7}. \REV{Values sampled from a Rayleigh distribution can in principle go above the physical upper limits of eccentricity and inclination ($1$ and $\pi$, respectively). We make sure that we discard and resample such values whenever they occur, which is extremely unlikely even for the set with $\sigma = 0.1$ (the probability density at $e=1$ for $\sigma_e = 0.1$ is $10^{-20}$).
}

These eight sets represent discs with increasing velocity dispersion, and can be interpreted as different evolutionary stages of the same disc, whose orbits become more eccentric and inclined due to mutual gravitational interactions. Our $\sigma$ parameter can be thought of as an analogue to the $W_0$ parameter of Michie-King globular clusters models \citep{michie1963,king1966}, which controls their central density, and for which higher values of $W_0$ correspond to more evolved clusters due to two-body relaxation \citep{spitzer1987}.

%-------------------------------------- Two column figure (place early!)
%

%--------------------------------------------------- One column table
\begin{table}
        \caption{Initial conditions for our sets of simulations.}
        \centering
        \begin{tabular}{l|c}
                Property &  Values \\
                \hline
                $\ainn$ & $\log{\cal U}(0.1, \Rhill/2)$ \\
                $\einn$ &  0\\
                $\iinn$ &  0\\
                $\esin, \eout$ & ${\cal R}(\sigma)$ \\
                
                $\isin$ & ${\cal R}(\sigma/2)$ \\
                $m_1, m_2, m_{\rm sin}$ & $\log{\cal U}(10, 50) \,\msun$ \\
                $\log_{10}(\sigma)$ & $-1, -2, ..., -7, -\infty$ \\
        \end{tabular}
        \tablefoot{Initial conditions for our sets of simulations. $\ainn$, $\einn$: semi-major axis and eccentricity of the (inner) BH binary orbit. $\iinn$: inclination of the inner binary orbit with respect to its orbit around the SMBH. $\esin, \eout$: eccentricities of the binary and the single around the SMBH. $\isin$: inclination of the single with respect to the binary orbit around the SMBH. ${\cal R}(\sigma)$ denotes a Rayleigh distribution with scale parameter $\sigma$. $\log{\cal U}(a,b)$ indicates a log-uniform distribution between $a$ and $b$.
        }
        \label{tab:ic}
\end{table}

We assume the binary inner orbit to be initially circular ($\einn=0$). As the  outer orbit of the binary defines a plane and therefore the direction of the orbital angular momentum of the binary relative to the SMBH, here we assume the inner orbit to be prograde and coplanar.\footnote{In this dry BH nucleus, we neglect orbital precession of the orbits as it occurs on a timescale longer than the dynamical interaction time. We therefore disregard the spin of the SMBH.}
The binary semi-major axis is drawn from a log-uniform distribution ($\log{\cal U}$) between 0.1 $\au$ and half of its Hill radius $\Rhill$, which is given by
\begin{equation}\label{eq:Rhill}
        \Rhill = \frac{1}{2} \aout (1 - \eout) \left( \frac{\mout}{3\msmbh} \right)^{1/3} \:,
\end{equation}
where $\mout=m_1 + m_2$ is the total mass of the binary and $\msmbh$ is the mass of the SMBH. We sample the mass of the stellar-mass BHs from a log-uniform distribution between 10 and $50 \,\msun$ and set the mass of the SMBH to $4.31 \times 10^{6} \,\msun$, which is the inferred mass of SgrA* \citep{gil09a,gil17}. The true anomaly of the inner BH binary is uniformly sampled between 0 and $2\pi$.

The binary and the single orbits cross at a distance $\Renc$ from the SMBH. We set this distance by drawing it from a $f(\Renc) \propto \Renc$ distribution between $0.03 \,\pc$ and $0.1 \,\pc$; this latter range is inspired by the stellar disc observed around SgrA*. Given $\Renc$, $\eout,$ and $\esin$, we sample $\asin$ and $\aout$ from the same $\Renc$ distribution, adding the constraint that the orbits need to cross at $\Renc$. The orbit of the binary is fixed so that it passes through $\Renc$, and we add an impact parameter with respect to this location in order to determine where the orbit of the single will pass. The impact parameter is set to be orthogonal to the tangential velocity of the single at the encounter location, and we sample its magnitude from a $f(b) \propto b$ distribution between $0$ and $2 \ainn$.

Once $\Renc$, $\esin$, $\eout$, $\asin$, $\aout$, and $\isin$ are fixed, the orbital parameters of the single and the centre of mass of the binary are uniquely determined. We then start the simulations with the binary and the single some time $\Delta T_{\rm enc}$ before the encounter along their orbits. Here, we chose $\Delta T_{\rm enc}=\min(P_{\rm out},P_{\rm sin}) / 16$, where $P_{\rm out}$ and $P_{\rm sin}$ are the orbital periods of the binary and the single around the SMBH. We increase $\Delta T_{\rm enc}$ if the binary and the single are already within the sum of their Hill radii at the beginning of the simulations, or if they are closer than 100 times the binary semi-major axis. In set~\sszero{}, the orbits are perfectly circular and aligned, meaning that the impact parameter is also the difference in semi-major axis between the two orbits; for more details, see \citet{trani2019a,trani2019b}. We run a total of $5\times 10^4$ realisations per set. The initial conditions are summarised in Table~\ref{tab:ic}.

\subsection{The \textsc{tsunami} code}

We run the simulations with \textsc{tsunami}, an implementation of Mikkola's algorithmic regularisation \citep{mik99a,mik99b,trani2023iaus}.  \textsc{tsunami} is ideally suited for integrating the dynamical evolution of few-body systems in which strong gravitational encounters are frequent and the mass ratio between the interacting bodies is large. It implements the logarithmic Hamiltonian and the time-transformed leapfrog, along with the non-regularised leapfrog as described in \citet{mikkola2020}. In the present paper, we employ the logarithmic Hamiltonian scheme. \textsc{tsunami} uses a Bulirsch-Stoer algorithm with stepsize control to improve the accuracy of the integration, which would otherwise be of second order \citep{stoer1980}. Finally, \textsc{tsunami} solves the equations of motion in a system of relative coordinates based on a chain of inter-particle vectors in order to reduce round-off errors when calculating small distances between bodies \citep{mik93}. \textsc{tsunami} includes velocity-dependent forces, such as post-Newtonian (PN) corrections and tidal forces \citep{mikkola2006,mikkola2013}. In this work, we enable PN terms of order 1PN, 2PN, 2.5PN, and 3.5PN \citep{blanchet2014}.

\subsection{Stopping criteria}

The initial time of the simulation is defined when the single comes within $2\ainn$  of the centre of mass of the binary.
Since exchanges and ionisations can occur during the simulation, we identify the particles forming the bound binary at any given moment in time. Specifically, we check for the most bound stellar-mass pair in the system and label it as the binary, while the remaining body is labelled as the single. We flag an interaction as terminated if the single is unbound and has positive radial velocity with respect to the binary. We then terminate the simulation once the binary--single distance exceeds 20 binary semi-major axes. We also stop a simulation when there is no bound binary, and all stellar-mass bodies are unbound with respect to each other and their common centre of mass. Finally, we define a merger when two bodies become closer than the sum of their collision radii, which we set to ten times their Schwarzschild radius.

\subsection{Classification of interactions}\label{sec:class}

A three-body interaction can be thought of as a succession of two micro-states: (1) `democratic interactions' where all three bodies are close to each other and freely exchange angular momentum and energy, and (2) `excursions', where the binary and the single recoil on a bound orbit, temporarily forming an unstable hierarchical triple \citep[e.g.][]{mon76a,mon76b,valtonen2005,stone2019,kol2021,ginat2021}. Eventually, in Newtonian gravity, an unstable three-body system will inevitably decay into an unbound binary--single pair; that is, the single escapes the system.

\begin{figure}
        \centering
        \includegraphics[width=\columnwidth,trim=2.5cm 1.2cm 1.3cm 2.3cm,clip]{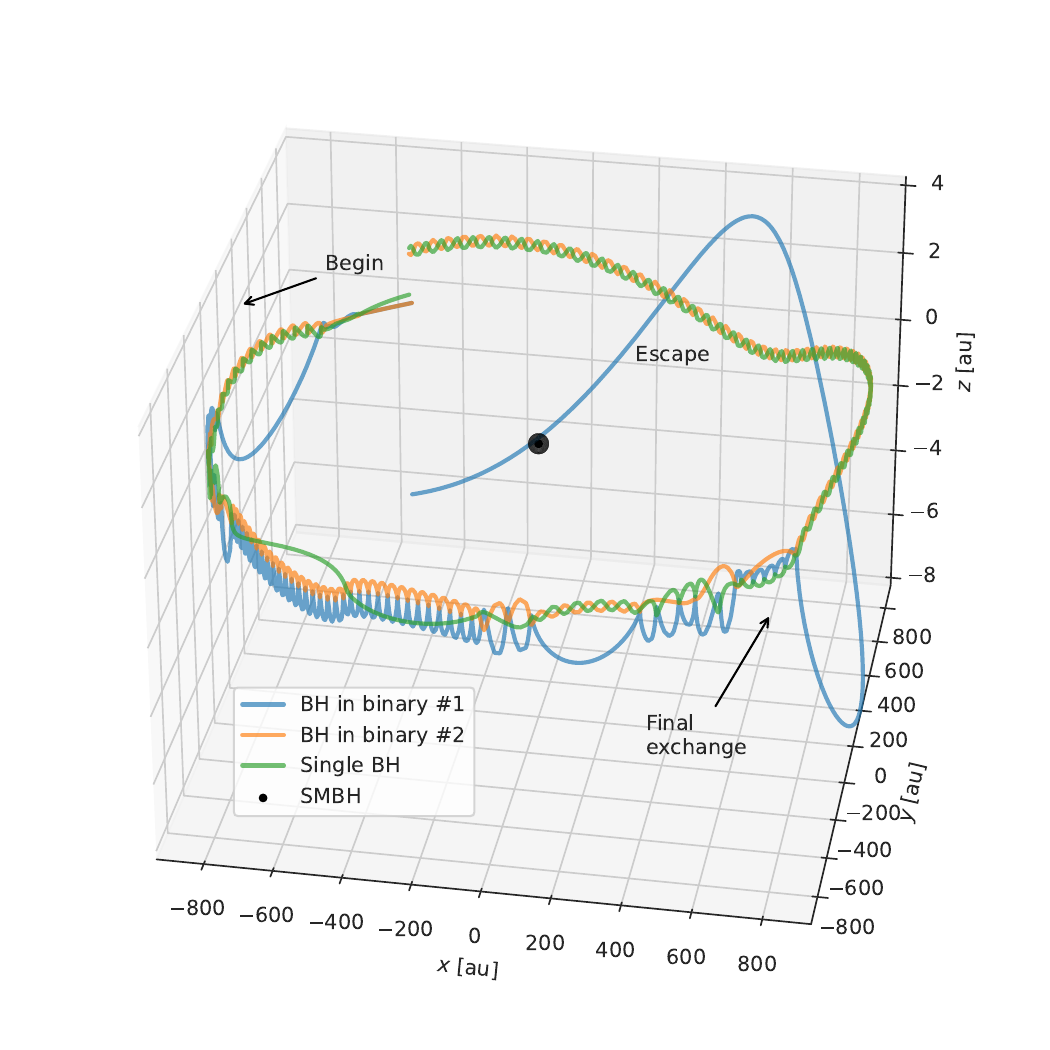}
        \caption{Trajectories of the three stellar-mass BHs during a three-body encounter around the SMBH. We highlight the different scale of the $z$ axis with respect to the $x$-$y$ coordinates, used in order to better visualise the trajectories. The interaction is resonant, because there are multiple episodes of democratic interactions and excursions.}
        \label{fig:traj}%
\end{figure}

Throughout the paper, we employ the following language to describe the qualitative nature of a three-body interaction.
If the escape of the single happens after only one democratic interaction without any excursions, we label the interaction as `prompt'. On the other hand, if there is at least one excursion, we label the interaction as `resonant' (using the terminology established in \citealt{hut83a}). A flyby, where a single passes close to the binary without breaking it and escapes immediately, is necessarily a prompt interaction. On the other hand, an exchange, where the single takes the place of a binary member, can result from both prompt and resonant interactions \citep[see e.g.][]{manwadkar2020,manwadkar2021}. Analogously, if the final binary is the same as the initial one, it might have resulted from a prompt interaction ---a flyby--- or from a resonant interaction ---possibly after several exchanges with the single. For this reason, we label those binaries formed by the same bodies that made up the initial binary as `original', regardless of whether they were formed via a resonant or a prompt interaction. Figure~\ref{fig:traj} displays the trajectories of the stellar-mass BHs during a resonant exchange.

\section{Results and discussion}

\begin{figure}
        \centering
        \includegraphics[width=\columnwidth]{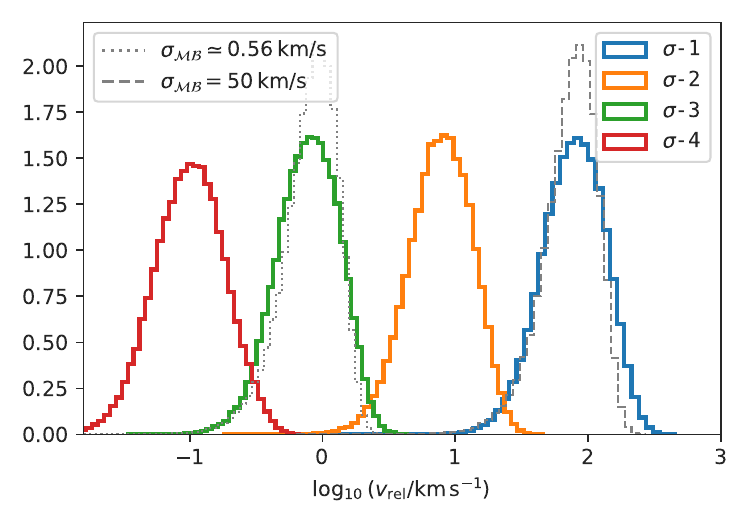}
        \caption{Distributions of relative velocity between the centre of mass of the binary and that of the single at the encounter location $\Renc$, for different disc velocity dispersions $\sigma$. The blue, orange, and red histograms correspond to $\sigma=10^{-1}, 10^{-2} , 10^{-3}$, and $10^{-4}$, respectively. The grey dotted line is a Maxwell-Boltzmann distribution, which is characteristic of isotropic clusters, and matches the velocity dispersion in \sset{-2}. The grey dashed line indicates the typical velocity distribution for a NSC that does not host a central SMBH.
        }
        \label{fig:venc}%
\end{figure}

\subsection{Varying disc velocity dispersions}\label{sec:vdisc}
As explained in Section~\ref{sec:ics}, the disc velocity dispersion $\sigma$ controls the relative velocity ($v_{\rm rel}$) between the binary and the single at the encounter location, which can be thought of as the velocity at infinity in a binary--single encounter in isolation, before the encounter. To better understand the impact of the disc velocity dispersion, we show in Figure~\ref{fig:venc} the distribution of $v_{\rm rel}$ corresponding to $\sigma = 10^{-1}, 10^{-2}, 10^{-3}$, and $10^{-4}$. For lower disc velocity dispersion, the difference in velocity only arises from the Keplerian shear, and is of the order of $10^{-2}\,\rm km\,s^{-1}$. For comparison, we also show two Maxwell-Boltzmann distributions with a dispersion of $\sigma_{\cal MB}$, which accurately describe the velocity dispersion in an isotropic cluster. 

%--------------------------------------------------- One column table
\begin{table*}
        \centering
        \caption{Outcomes of the simulations by percentage, following the classification scheme explained in Section~\ref{sec:class}.}
        \begin{tabular}{l|rrrrrrrr}
                Set & \sset{-1}            & \sset{-2}          &  \sset{-3}  & \sset{-4}               & \sset{-5}          &  \sset{-6} &  \sset{-7} &  \sszero \\ \hline\hline \noalign{\vskip 3pt}
Mean $v_{\rm rel}$ [$\rm km/s$] & 55.26 &5.53 & 0.56 & 0.07 & 0.04 & 0.04 & 0.04 & 0.04 \\
Hard binaries [\%] & 73.23 & ${\sim}$100 & 100 & 100 & 100 & 100 & 100 & 100 \\\hline \noalign{\vskip 3pt}
Prompt flyby & 11.58 & 7.72 & 7.63 & 7.72 & 7.80 & 7.79 & 7.49 & 7.65 \\
Prompt exchange & 46.07 & 33.51 & 34.89 & 34.21 & 32.93 & 33.26 & 33.49 & 33.24 \\
Resonant original & 18.48 & 25.66 & 23.59 & 21.12 & 18.65 & 17.01 & 16.78 & 16.56 \\
Resonant exchange & 18.60 & 32.59 & 32.83 & 31.23 & 27.68 & 26.24 & 25.78 & 25.63 \\
Breakup & 4.87 & 0.00 & 0.00 & 0.00 & 0.00 & 0.00 & 0.00 & 0.00 \\
Merger & 0.39 & 0.52 & 1.06 & 5.72 & 12.95 & 15.70 & 16.47 & 16.92 \\
$t_\mathrm{GW} < t_\mathrm{Hubble}$ & 35.44 & 37.14 & 53.29 & 60.16 & 57.54 & 56.13 & 55.85 & 55.34 \\
\multirow{2}*{$e > 0.1$ at $f_\mathrm{GW} = 10 \rm\,Hz$}  & \multirow{2}*{0.16}  & \multirow{2}*{0.21} & \multirow{2}*{0.44} & \multirow{2}*{2.56} & \multirow{2}*{6.18} & \multirow{2}*{7.35} & \multirow{2}*{7.75} & \multirow{2}*{8.01}\\
& & & & & & & & \\
"" over in-cluster mergers & 40.15 &40.79 & 41.74 & 44.80 & 47.92 & 46.95 & 47.18 & 47.43 \\
                \hline\noalign{\vskip 3pt}
        \end{tabular} 
		\tablefoot{
	         	   The third-to-last row shows the percentage of merging binaries with  a GW coalescence timescale ($t_\mathrm{GW}$) of the final binary of less than a Hubble time ($t_\mathrm{Hubble} \equiv 1.4 \times 10^{10}\rm\,yr$). The second-to-last and the last rows show the percentage of highly eccentric mergers ($e > 0.1$) at $f_\mathrm{GW} = 10 \rm\,Hz$ in proportion to the total number of interactions and in-cluster mergers, respectively. \REV{The first two rows indicate the initial mean relative velocity between the binary--single and binary fraction, respectively.}
        }
        \label{tab:outcomes}
\end{table*}

The velocity dispersion in a Keplerian disc is qualitatively different from that of an isotropic star cluster. A Maxwell-Boltzmann distribution with $\sigma_{\cal MB} = 0.6 \rm\, km\,s^{-1}$ is too peaked and sharp to efficiently approximate the $v_{\rm rel}$ distribution of the $\sigma=10^{-2}$ set. Most importantly, $\sigma_{\cal MB} = 0.6 \rm\, km\,s^{-1}$ is typically associated with low-mass open clusters and not with a nuclear star cluster (NSC). A typical NSC, like that of the Milky Way, would have a velocity dispersion of $\sigma_{\cal MB} = 50 \rm\, km\,s^{-1}$ if it did not have an SMBH at its centre \citep[e.g.][]{atallah2022}, which is comparable only with the most eccentric and misaligned disc we consider ($\sigma = 10^{-1}, e, i \sim 0.125, 3.6^\circ$). 

The low velocity dispersion in a Keplerian disc has major implications for the outcome of three-body encounters. Binaries are more likely to be hard in a low-velocity-dispersion environment, which makes them more likely to harden, that is, to shrink in separation \citep{heg75}. The chance that a binary is hard is also increased by the dynamical stability constraint of Equation~\ref{eq:Rhill}, which enforces a strict upper limit on the size of binaries, beyond which they will be disrupted by the tidal field of the SMBH. \REV{Given our choice of initial conditions, we find that only 73.22\% of the binaries in set~\sset{-1} have a binding energy greater than the kinetic energy of the incoming single. In every other set, all binaries are hard.} This explains why set~\sset{-1}  displays qualitatively different outcomes with respect to sets~\sset{-2} and \sset{-3}, as we show below.

%----------------------------------------------------------------- 
\begin{figure}
        \centering
        \includegraphics[width=\columnwidth]{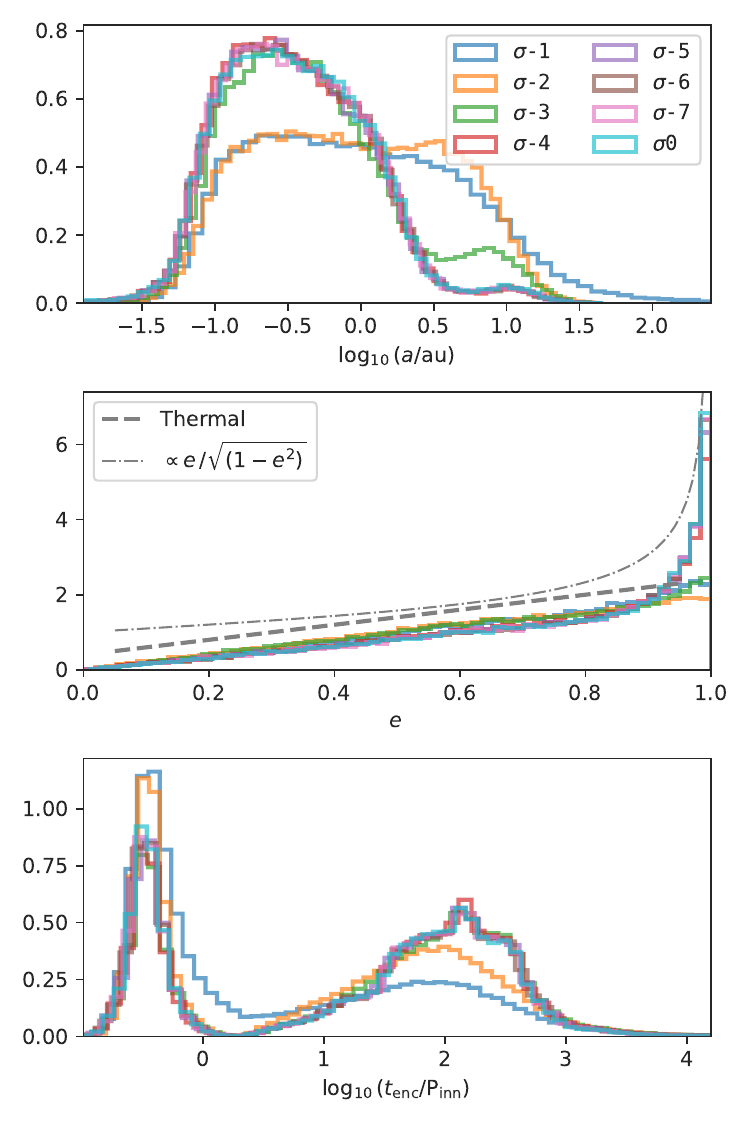}
        \caption{Distributions of binary semi-major axes (top panel) and eccentricities (middle panel) of post-encounter binaries, and encounter duration (bottom panel) for different disc velocity dispersions $\sigma$. The blue, orange, and green histograms correspond to $\sigma=10^{-1}, 10^{-2}$, and $10^{-3}$, respectively. The encounter duration is shown in units of the initial inner binary period, which is different for each realisation. The eccentricity distributions become more skewed towards $e\sim1$ as the disc velocity dispersion decreases, so that for $\sigma \geq 10^{-3}$ they are roughly thermal, while for $\sigma \leq 10^{-4}$ they are superthermal. The distribution of encounter duration shows two peaks, corresponding to prompt and resonant interactions, with the former peak being more pronounced in set~\sset{-1}.
        }
        \label{fig:aeenc}%
\end{figure}

%-----------------------------------------------------------------

For decreasing velocity dispersion, the centre of mass of the binary  is close to that of the single for a longer period of time before the encounter, which makes them interact gravitationally with each other even before the encounter happens. This breaks our assumption that the two bodies keep on their original Keplerian before the encounter, and causes them to miss each other  upon a number of encounters; that is, the single never comes within $2 \ainn$ of the binary. We exclude these flybys from the following statistical analysis because the encounter does not actually happen. The frequency of such cases is negligible for \sset{-1} and \sset{-2}, but it increases to ${\sim}34\%$ for \sset{-3} and to ${\sim}38\%$ for $\sigma \leq 10^{-4}$.

Table~\ref{tab:outcomes} summarises the simulation outcomes. \REV{The only set that presents breakups of the binary is set~\sset{-1}, which is consistent with it being the one with 27\% of the binaries being soft.} Both the number of mergers during the simulations and the number of binaries with a GW coalescence time ($t_\mathrm{GW}$) of less than a Hubble time ($t_\mathrm{Hubble}\equiv 14 \rm\, Gyr$) increase with decreasing velocity dispersion. Similarly, the number of prompt interactions is higher in \sset{-1}, as expected from its high velocity dispersion. All the mergers happen between the stellar-mass BHs, and do not involve the SMBH. In principle, the escaping single might be kicked onto a highly eccentric orbit around the SMBH, triggering an extreme-mass-ratio inspiral \citep{amaro-seoane2007}. We calculated the post-encounter orbital properties of the single and estimated its coalescence time with the SMBH, but we find no inspirals that would occur within $t_\mathrm{Hubble}$.

Interestingly, the fraction of resonant interactions changes in a non-trivial way as a function of the velocity dispersion. The highest percentage (${\sim}59\%$) of resonant encounters occurs at $\sigma = 10^{-2}$ and decreases for lower velocity dispersions to ${\sim}54\%$ at $\sigma=0$ (we note that Table~\ref{tab:outcomes} shows this classification only for encounters with a post-encounter binary).

The top two panels of Figure~\ref{fig:aeenc} show the final binary properties for those simulations that do not end in a breakup or merger. The semi-major axes tend to be larger for \sset{-1}, whose binaries in the upper tail of the distribution are close to being disrupted by the tidal field of the SMBH. All the sets with $\sigma \leq 10^{-4}$ do not display significant differences in the semi-major axes distributions.

The eccentricity distribution of the binaries after the encounter is also strongly dependent on the disc velocity dispersion. For $\sigma \leq 10^{-4}$, the eccentricity distribution is remarkably superthermal, with an excess of eccentric binaries at $e \sim 1$. The superthermal eccentricity of low-angular momentum triples has often been described by the distribution $f(e) \propto e / \sqrt{(1 - e^2)}$ \citep[e.g.][]{valtonen2005}, but in our case it is rather a mix of thermal eccentricities at $e \lesssim 0.9$ and a superthermal component at $e \gtrsim 0.9$.
For $\sigma \geq 10^{-3}$, the eccentricity distribution is very close to a thermal distribution ($f(e) \propto e$), with just a slight excess of highly eccentric binaries at $e\sim1$, which is more prominent in \sset{-3}. 

\REVV{This result can be explained in terms of the geometry of the  system. As inclination and eccentricity decrease, the binary--single interaction retains less angular momentum and becomes increasingly confined to a 2D plane. This restriction to the phase-space dimensionality of the interacting triple leads to higher eccentricities in the final binary. Although this never reaches the extremes observed in systems with zero angular momentum or those fully confined to two dimensions \citep[as in the equilateral triangle experiments of][]{hugopaper}, the trend is evident, and one can observe the eccentricities shifting to a superthermal distribution.} The eccentricity has a strong impact on the GW coalescence timescale \citep{peters64}, which makes the number of binaries with $t_\mathrm{GW} < t_\mathrm{Hubble}$ decrease with increasing velocity dispersion.

The bottom plot in Figure~\ref{fig:aeenc} shows the distribution of encounter durations in units of the initial binary period. All the distributions are clearly bimodal, with the left and right peaks corresponding to prompt and resonant interactions, respectively.

\REV{
Overall, these results agree with what we would expect from basic considerations. The disc velocity dispersion affects the eccentricity and inclination of orbits, which in turn determine the relative velocity between the binary and the single. An aligned and coplanar disc will have the lowest relative velocity during encounters, which will lead to highly eccentric binaries and more GW mergers.

However, upon closer examination of the merger fraction concerning velocity dispersion, a notable trend emerges. The merger fraction rises more rapidly as the velocity dispersion decreases than anticipated by analytic expectations. \citet{ginat2023nonth} establish semi-analytically that the merger probability ---as a function of the total energy of the triple--- is well approximated by a power law with an exponent of  ${\approx} 3.5$, as illustrated in their Figure~1. Repeating a parallel analysis using our simulated data, we observe that the merger fraction is only broadly consistent with a power-law exponent of ${\approx} 4.0$. Moreover, it exhibits a more pronounced increase at low energies and a more gradual decrease at higher energies compared to a power-law trend. The observed discrepancy is ascribed to the tidal field of the SMBH, which introduces qualitative changes in the behaviour of dynamical interactions. This aspect is further investigated in the following section.
}

\subsection{The role of the tidal field of the SMBH}\label{sec:smbhtid}

%--------------------------------------------------- One column table
\begin{table*}
		\caption{Outcomes of the simulations in isolation by percentage, following the classification scheme explained in Section~\ref{sec:class}.}
        \centering
        \begin{tabular}{l|rrrrrrrrr}
        Set & \sset{-{$\cal MB$}} & \sset{-1-i}            & \sset{-2-i}                &  \sset{-3-i}  & \sset{-4-i}                & \sset{-5-i}                &  \sset{-6-i} &  \sset{-7-i} &  \texttt{$\sigma$0-i} \\ \hline\hline \noalign{\vskip 3pt}
Mean $v_{\rm rel}$ [$\rm km/s$] & 50.05 & 55.26 &5.53 & 0.56 & 0.07 & 0.04 & 0.04 & 0.04 & 0.04 \\ 
Hard binaries [\%] & 75.57 & 73.23 & ${\sim}$100 & 100 & 100 & 100 & 100 & 100 & 100 \\\hline \noalign{\vskip 3pt}
Prompt flyby & 20.57 & 16.41 & 3.54 & 4.37 & 6.58 & 11.89 & 11.96 & 12.16 & 11.99 \\
Prompt exchange & 38.81 & 48.36 & 46.32 & 44.88 & 43.04 & 38.55 & 38.15 & 38.27 & 38.30 \\
Resonant original & 18.74 & 15.42 & 20.68 & 19.36 & 18.58 & 18.31 & 18.31 & 18.31 & 18.23 \\
Resonant exchange & 19.39 & 14.43 & 23.09 & 22.26 & 22.28 & 23.40 & 23.39 & 23.11 & 23.15 \\
Breakup & 2.29 & 3.84 & 0.00 & 0.00 & 0.00 & 0.00 & 0.00 & 0.00 & 0.00 \\
Merger & 0.22 & 1.58 & 6.39 & 9.14 & 9.54 & 7.86 & 8.20 & 8.17 & 8.34 \\
$t_\mathrm{GW} < t_\mathrm{Hubble}$ & 35.52 & 41.55 & 44.92 & 44.09 & 42.31 & 34.63 & 34.30 & 34.30 & 34.09 \\
\multirow{2}*{$e > 0.1$ at $f_\mathrm{GW} = 10 \rm\,Hz$}  & \multirow{2}*{0.11}  & \multirow{2}*{0.75} & \multirow{2}*{2.72} & \multirow{2}*{3.50} & \multirow{2}*{3.98} & \multirow{2}*{3.42} & \multirow{2}*{3.58} & \multirow{2}*{3.50} & \multirow{2}*{3.55} \\
& & & & & & & & & \\
"" over in-cluster mergers & 50.00 &47.47 & 42.56 & 38.39 & 41.90 & 43.56 & 43.78 & 42.94 & 42.70 \\
        \hline\noalign{\vskip 3pt}
\end{tabular}      
        \tablefoot{
            The simulations sets with the \texttt{-i} suffix have the same velocity dispersion as the ones in Table~\ref{tab:outcomes}, but the encounter happens in isolation (i.e. without the SMBH's tidal field). For \sset{-{$\cal MB$}} the velocity dispersion is given by a Maxwell-Boltzmann distribution with $\sigma_{\cal MB} = 50 \rm\, km\,s^{-1}$ and the binaries are randomly oriented relative to the incoming single. The rows are the same as in Table~\ref{tab:outcomes}.
        }
        \label{tab:smbh_iso}
\end{table*}

\subsubsection{Expectations from analytic theories and previous results}

Our simulations fully include the effect of the tidal field of the SMBH, which is modelled as a point particle. Based on the results of \citet{trani2019b}, we can expect the tidal field to have two main effects on the three-body dynamics. 
First, the  encounter will be shorter in duration than  encounters in isolation because the system is tidally limited. Consequently, a three-body interaction may end prematurely with the first excursion that extends beyond the tidal radius of the system. Because the interaction would then go through fewer democratic interactions, we can expect the final binaries to be less eccentric and softer than if the same encounter had happened in isolation. 
Second, the tidal field will exert a torque on the three-body system, which will make the system gain or lose angular momentum. Low-angular-momentum triples are the ones that result in a superthermal distribution of final binaries \citep[e.g.][]{mik86,valt05,manwadkar2023,leigh2022therm,hugopaper,ginat2023nonth}, and therefore have the highest chance of producing a GW coalescence. \citet{trani2019b} found that low-angular-momentum triples in a Keplerian tidal field will tend to gain angular momentum, so that the final binary eccentricity distribution will be thermal rather than superthermal.

It is easy to estimate the truncating effect of the tidal field using simple calculations. As shown in \citet{trani2019b}, the tidal field of the binary limits how extended the outer orbit of the temporarily bound triple  can be during an excursion. This limit is represented by the Hill radius, $\Rhill$, of the triple system. From energy conservation, this implies that the upper limit of the semi-major axis that the inner binary can have increases for decreasing $\Rhill$:
\begin{equation}\label{eq:amax}
        a^{\rm max} =  -\frac{1}{\frac{2 E_0}{G m_1 m_2} + \frac{m m_3}{m_1 m_2}\frac{1}{\Rhill}}\:,
\end{equation}
where $E_0$ is the total energy, $m = m_1 + m_2$ is the inner binary mass, and $m_3$ is the tertiary body (see also equation~6 of \citealt{trani2019b}). This result was confirmed by numerical experiments and later extended by \citet{ginat2021pot} who derived the resulting distribution of the binary semi-major axes from statistical escape theories using an arbitrary distance cut-off instead of $\Rhill$ (see their Figure~1). 

Based on these results, we expect a shift in the distribution of semi-major axes towards larger values as the effect of the tidal field is stronger. In our simulations, all the sets have comparable $\Rhill$ values, which range from ${\sim}50\rm\,\au$ to ${\sim}200 \rm\,\au$, with a median value of ${\sim}134 \rm\,\au$.  We can then estimate  the strength of the tidal field by comparing Equation~\ref{eq:amax} ---using the values of $E_0$ and $\Rhill$ from the simulations--- with the same expression for $\Rhill \rightarrow +\infty$. Even though the distribution of $a^{\rm max}$ shifts to smaller values compared to that of $\lim_{\Rhill \rightarrow +\infty} a^{\rm max}$, both distributions have similar median values of $a^{\rm max} \simeq 0.6 \au$. 

This suggests that the effect of the tidal field will be relatively small, given our set of initial conditions. However, it is worth noting that the estimate above only considers the truncating effect of the tidal field, neglecting the impact due to exchanges of energy and angular momentum with the triple system. 
Instead of investigating the impact of varying tidal field strengths on the outcomes of three-body interactions, which was previously addressed in \citet{trani2019b}, this section focuses on a different comparison. In the following subsection, we introduce a new set of simulations specifically designed to isolate the influence of the tidal field. We compare three-body interactions occurring in the vicinity of the SMBH with those taking place in an isolated environment.

\begin{figure}[h]
        \centering
        \includegraphics[width=\columnwidth]{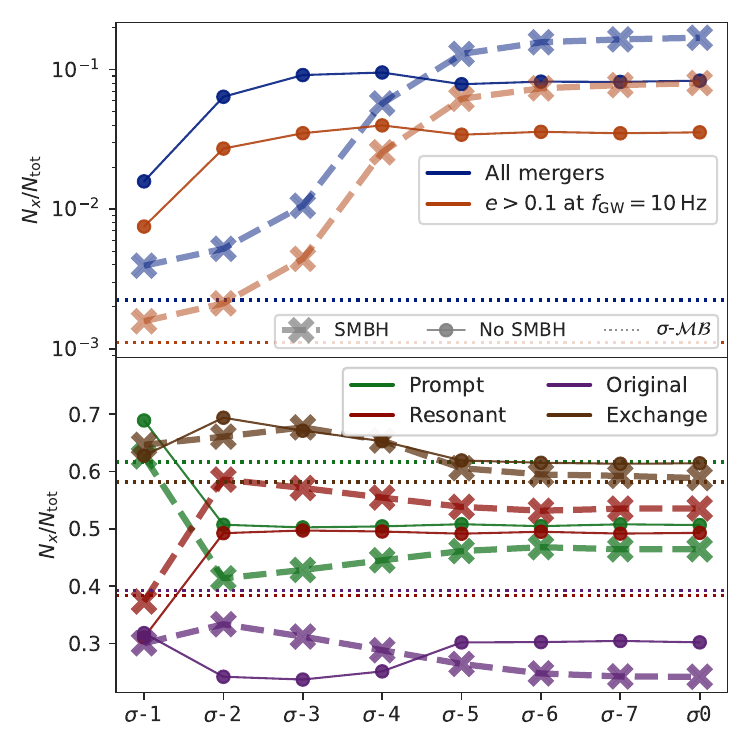}
        \caption{
                Fraction of outcomes for each set of simulations. The velocity dispersion decreases from left to right, and goes from $\sigma = 10^{-1}$ to $\sigma = 0$. The solid lines with circle markers are the simulation in isolation ($\sigma n \texttt{-i}$). The dashed lines with cross markers indicate the simulations with the SMBH ($\sigma n$). The horizontal dotted lines indicate the results for \sset{-${\cal MB}$}. 
                The top panel shows the fraction of mergers (blue lines) and the fraction of mergers with high eccentricity at $f_{\rm GW} = 10 \,\rm Hz$ (orange lines). The bottom panel shows the fraction of prompt (green lines) and resonant (red lines) interactions, and the fraction of encounters ending with the original binary (purple lines) or with an exchanged member (brown lines).
                %Top panel: cumulative distribution of GW inspiral times ($t_{\rm GW}$) for sets~\sset{-2} (rose), \sset{-2-i} (fawn) and \sset{-2-{$\cal MB$}} (teal). The simulations without the SMBH's tidal field display a strong enhancement at small $t_{\rm GW}$, due to the presence of many eccentric binaries.
                %Middle panel: distribution of binary eccentricities at GW peak frequency $f_{\rm GW} = 10\rm\,Hz$ for the mergers occurring during the simulations. The histograms of the main panel are normalized to the total number of mergers. The inset panel shows the distribution of $\log{e}$ of each set, normalized to 1. Including the effect of the SMBH's tidal field causes the number of eccentric mergers to drop by a factor of ${\sim}24$.
                %Bottom panel: distribution of spin-orbit tilt angle ($\theta$) caused by the tilting of the binary plane during the encounter. An isotropic distribution would appear as flat in $\cos{\theta}$. In the presence of the SMBH's tidal field, the bimodality of the tilt angle distribution disappears, leaving only spin-orbit aligned binaries.
        }
        \label{fig:outcomp}%
\end{figure}

\begin{figure*}
        \centering
        \includegraphics[width=\linewidth,trim={0 0.3cm 0 0},clip]{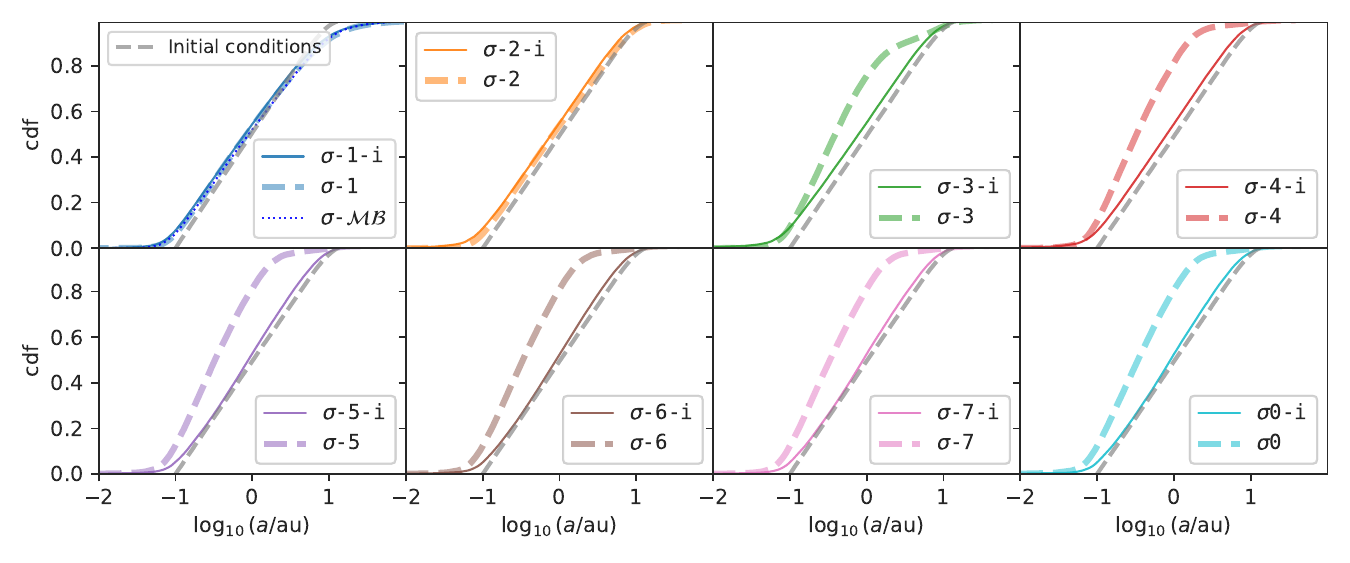}
	
        \includegraphics[width=\linewidth,trim={0 0 0 0.3cm},clip]{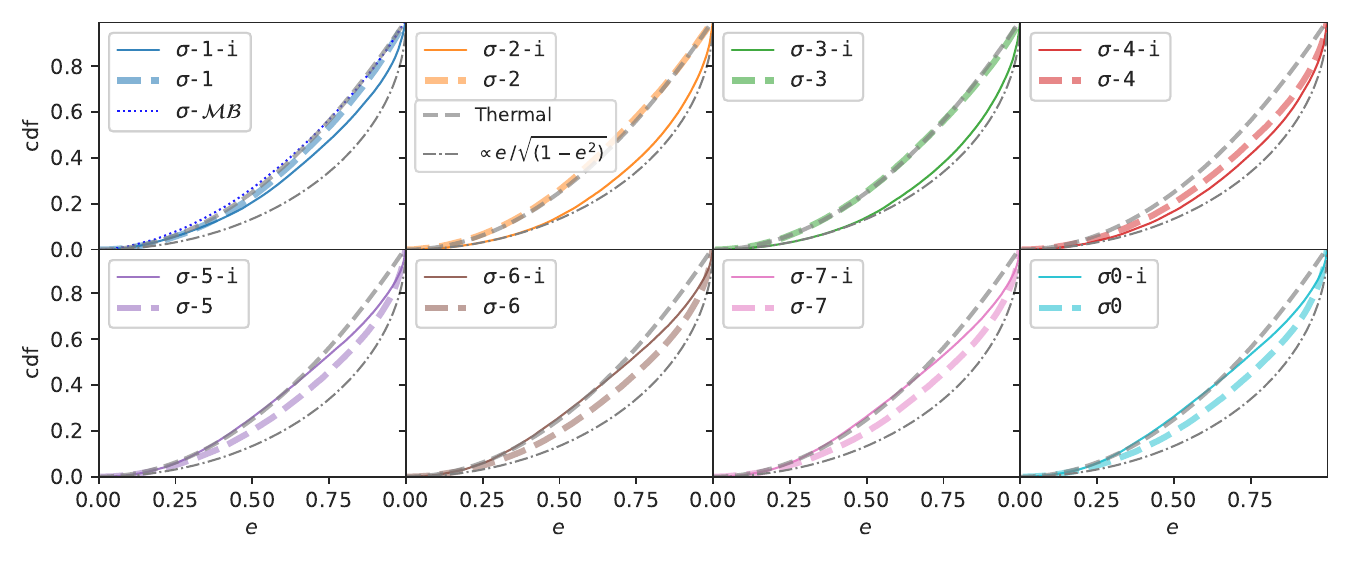}
        \caption{Cumulative distributions of binary semi-major axes (top panels) and eccentricities (bottom panels) of post-encounter binaries. Solid thin lines indicate the sets with the SMBH (\sset{n}), while the thick dashed line indicates the sets in isolation (\sset{n-i}). The disc velocity dispersion for each set of curves decreases from left to right and from top to bottom. For $\sigma \geq 10^{-4}$, the number of highly eccentric binaries is higher in the absence of tidal field. For $\sigma \leq 10^{-5}$, the situation is the opposite, and encounters around the SMBH result in slightly higher eccentricities.
        }
        \label{fig:smbh_iso}%
\end{figure*}

\subsubsection{Comparison with isolated encounters}

To quantify the role of the tidal field, we performed simulations without the SMBH using the relative velocity (Figure~\ref{fig:venc}) derived from the simulations with the SMBH as the velocity at infinity in a hyperbolic binary--single encounter in isolation. For the other initial conditions, we used exactly the same properties (masses, inclination of the single, etc.) as those in the sets with the SMBH. We indicate these sets with the suffix $\texttt{-i}$ to specify that the three-body encounter occurred in isolation; for example $\sigma\texttt{$n$-i}$, where $n$ goes from $-7$ to $0$. In this way, we still use the characteristic velocity dispersion associated to a Keplerian disc, but isolate the effect of the SMBH's tidal field. In these simulations, the inclination between the binary and the hyperbolic orbit is the same as $\isin$ in the SMBH simulations. Our setup is therefore similar to the that considered in \citet{samsing2022}, although we introduce a velocity dispersion characteristic similar to that found in a disc, rather than assuming that the bodies are initially at rest. 

Sets~\texttt{$\sigma n$} and \texttt{$\sigma n$-i} are meant to isolate the effect of the SMBH's tidal field in the same environmental conditions, that is, they both have a velocity dispersion associated with a Keplerian disc. However, we can also compare them to the environmental condition proper of a spherically symmetric, isotropic NSC without a central SMBH. Therefore, we performed an additional set of simulations, named \sset{-{$\cal MB$}}, with the same initial conditions as sets~\sset{-1-i}, but drawing the velocity at infinity from a Maxwell-Boltzmann distribution with $\sigma_{\cal MB} = 50 \rm\, km\,s^{-1}$. However, unlike in set~\sset{-1-i}, in \sset{-{$\cal MB,$}} the inclination between the binary and the hyperbolic orbit is drawn uniformly in $\cos{i}$, because in an NSC we expect no preferential direction.

Table~\ref{tab:smbh_iso} shows the frequency of final outcomes in the isolated sets, while Figure~\ref{fig:outcomp} compares the outcome fractions between the two sets. The number of resonant interactions becomes smaller in the absence of tidal field, and the number of prompt interactions increases accordingly. This outcome is both puzzling and unexpected because one would anticipate the tidal field to prematurely terminate the encounters, reducing the likelihood of the single body returning to interact with the binary. Consequently, one would expect a decrease in the frequency of resonant interactions. Instead, depending on the velocity dispersion, the sets without the SMBH exhibit between 8\% and 17\% less resonant interactions.

We find that this unexpected result is strictly tied to the definition of a resonant interaction, which we define as interactions with one or more excursions. If we use a stricter definition, allowing only interactions with two or more excursions, the results are reversed. In this case, the number of resonant interactions in isolation is from 8\% to 60\% larger compared to simulations with the SMBH, which agrees with our naive expectations. 

The source of this discrepancy lies in the fact that interactions around the SMBH have a significantly higher likelihood of experiencing one excursion compared to none, while the opposite is true for encounters in isolation. Thorough inspection and validation of the excursion-detection algorithm in our simulations revealed no issues.

We propose the following hypothesis to explain this behaviour. When the triple system disintegrates, the binary--single components are ejected on unbound orbits whether they are in isolation or around the SMBH. However, in the presence of the SMBH, the binary and the single continue on Keplerian orbits, which remain in close proximity unless the escape velocity of the binary--single is comparable to the orbital Keplerian velocity. Given the close proximity of the two orbits, they can more easily gravitationally focus and interact again, potentially leading to subsequent interactions. This contrasts with encounters in isolation, where the binary--single components remain unbound as they escape to infinity, precluding further gravitational interactions.

In practice, this mechanism requires that the escape velocity between the binary--single at breakup be sufficiently high that the two escape orbits become adequately separated. Otherwise, gravitational focusing may lead to their interaction. This can explain why interactions around the SMBH show a greater probability of having a single excursion compared to none, while the opposite is true for those in isolation.

Another way to see this mechanism is within the framework of the circular restricted three-body problem, where a test particle moves under the influence of two massive bodies, where one is significantly more massive than the other. If the test particle is within the Hill region of the less massive body, having negative binding energy with respect to it is not enough to guarantee escape; the particle may still remain trapped in the Hill region, `bouncing' against the zero-velocity surface before finding the Lagrange points, where the Hill region opens. 

Even though this mechanism exists, we find that the simulations in isolation have a longer duration with respect to those around the SMBH, confirming our initial hypothesis (and the findings of \citealt{trani2019b}) that the tidal field can interrupt excursions and end the encounters prematurely. The average encounter duration in isolation is between  approximately two and ten times longer than the average encounter duration of encounters around the SMBH, depending on the velocity dispersion.

More interesting is the evolution of the merger fraction for different velocity dispersions. 
In isolation, the merger fraction saturates at ${\sim}8\%$ for $\sigma \leq 10^{-5}$, while for the sets with the SMBH it rapidly increases in the same range.
Consequently, for $\sigma \geq 10^{-4}$, the sets in isolation have a consistently larger fraction of mergers, that is, up to ${\sim}12$ times larger at $\sigma = 10^{-2}$. This reverses for $\sigma \leq 10^{-5}$, for which the sets with the SMBH have a higher fraction of mergers, up to ${\sim}2$ times more at $\sigma = 0$.

The dependency on the disc velocity dispersion is further exemplified in Figure~\ref{fig:smbh_iso}, which compares the distributions of semi-major axis and eccentricity of the final binaries. 
The post-encounter binaries have smaller semi-major axes in the sets with the SMBH for $\sigma \leq 10^{-3}$.
The distribution of eccentricity also has a strong dependency on $\sigma$, which mimics the trend in the merger fraction. For $\sigma \geq 10^{-3}$, the eccentricity distribution for encounters with the SMBH is thermal, while it is superthermal for the sets in isolation. At $\sigma=10^{-4}$, the trend begins to reverse, with the distribution of \sset{-4-i} becoming more thermal and that of \sset{-4} becoming more superthermal. For $\sigma \leq 10^{-5}$, the eccentricity distribution is superthermal for both sets, although it is consistently more peaked at $e \sim 1$ in the sets with the SMBH.

This trend is consistent with both the results of \citet{trani2019b}, who find that the SMBH's tidal field causes the superthermal eccentricity distribution of binaries from low-angular momentum triples to become thermal, and those of \citet{samsing2022}, who instead find a superthermal eccentricity even without the presence of the SMBH. Here we show that the disc velocity dispersion is a critical parameter in determining the outcome of three-body encounters in discs, as it determines the shape of the final eccentricity distribution.

 The comparison between set~\sset{-$\cal MB$} and set~\sset{-1-i} is particularly intriguing. Despite both sets exhibiting similar velocity dispersions, as illustrated in Figure~\ref{fig:venc}, the binaries in set~\sset{-$\cal MB$} have an isotropic orientation, in contrast to the ordered orientation with $\langle i \rangle \sim \langle i\,\exp{(-{i^2}/{2 \sigma})}\rangle \sim 3.6^\circ$ observed in set~\sset{-1-i}. Consequently, the triple systems in set~\sset{-$\cal MB$} possess higher angular momentum overall, leading to a final eccentricity distribution of binaries that is thermal. This in turn brings it closer to set~\sset{-1}, which includes the SMBH. As a result, the merger fraction in set~\sset{-$\cal MB$} is significantly lower than in set~\sset{-1-i}. \REVV{This highlights the importance of the binary orientation with respect to the incoming single}.

In conclusion, including the presence of the SMBH is crucial to correctly estimate the outcome properties of three-body encounters around an SMBH. At high velocity dispersions ($\sigma \gtrsim 10^{-4}$), not including the SMBH results in overestimation of resonant encounters and mergers, while at low velocity dispersions it results in underestimation of the same quantities. 

\subsection{Implications for compact object mergers: Eccentricity and spin--orbit misalignment}\label{sec:gw}

\begin{figure*}
        \centering
        \includegraphics[width=\linewidth,trim={0 0.3cm 0 0},clip]{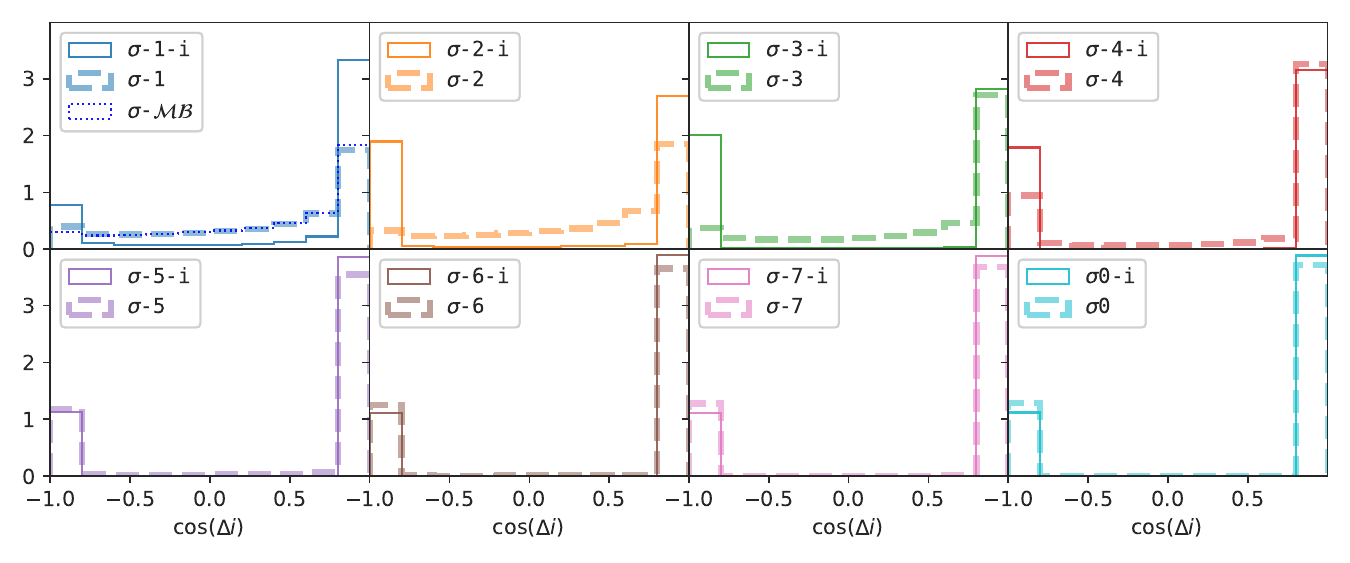}
        \caption{Distribution of spin--orbit tilt angle ($\Delta i$) caused by the tilting of the binary plane during the encounter. An isotropic distribution would appear as flat in $\cos{\theta}$. Solid thin lines indicate the sets with the SMBH (\sset{n}), while thick dashed lines indicate the sets in isolation (\sset{n-i}). The disc velocity dispersion for each distribution decreases from left to right and from top to bottom. For $\sigma \geq 10^{-4}$, the bimodality of the tilt angle distribution is strongly damped by the tidal field of the SMBH, leaving a majority of spin--orbit aligned binaries.
        }
        \label{fig:finincfig}%
\end{figure*}

Does the tidal field of the SMBH enhance or impair the formation of highly eccentric GW inspirals? As expected, the answer critically depends on the disc velocity dispersion.
We checked for eccentric inspirals that enter the LIGO-Virgo-Kagra band ($f_{\rm GW} = 10\rm\,Hz$) for the mergers that take place during the simulations by calculating the GW peak frequency ($f_{\rm GW}$) using the numerical fit of \citet{hamers2021fit}. The results are shown in the tables and in the top panel of Figure~\ref{fig:outcomp}.

The number of eccentric mergers closely traces the number of mergers that occur during the simulation, with about $40\%$ of the mergers retaining a significant eccentricity at $f_{\rm GW} = 10\rm\,Hz$. For $\sigma \gtrsim 10^{-4}$, the fraction of eccentric mergers is up to ${\sim}12$ times greater in the isolated sets than in the sets with the SMBH, and at $\sigma \leq 10^{-5}$ the fraction can be up to ${\sim}2$ times larger in the sets with the SMBH.
This is consistent with the results of the previous section, where we observe a transition in the eccentricity distributions at $\sigma = 10^{-4}$. 

In light of this, the simulations of \citet{samsing2022}, which were performed with zero velocity dispersion and neglect the SMBH, have likely underestimated the number of eccentric mergers by a factor of approximately two. However, the most significant comparison is with the set~\sset{-$\cal MB$}, which represents the initial conditions in NSC without an SMBH. The fraction of mergers, eccentric or not, is significantly smaller, from ${\sim}0.5$ times smaller at $\sigma = 10^{-1}$ to $1.4\times 10^{-2}$ smaller at $\sigma = 0$. Therefore, we can safely conclude that the outcome of three-body encounters in a BH disc around an SMBH strongly favours the production of highly eccentric mergers compared to encounters occurring in an NSC without an SMBH.

In addition to eccentricity, another property that can be used to discriminate GW formation channels is the spin--orbit misalignment. Depending on the formation pathways of binary BHs, their spin might be aligned with the total angular momentum of the orbit. This can be the case for binaries formed through gas capture in AGN discs, which are spun-up and aligned by gas accretion \citep{tagawa2020}. Another possible pathway to spin--orbit alignment is via the tidal spin-up of post-common-envelope binaries \citep[e.g.][]{kushnir2016,eldridge2016,kruckow2018,spera2019,bavera2020,belczynski2020,piran2020,tanikawa2020c}.

While it is generally assumed that binaries in dynamically active environments have isotropic spin--orbit misalignments, \citet{trani2021} showed that spin--orbit aligned BH binaries undergoing encounters retain memory of their initial configuration, resulting in non-isotropic tilt angle upon merger. \citet{samsing2022} found similar results for BHs undergoing encounters in an AGN disc.

The bottom panel of Figure~\ref{fig:finincfig} shows the distribution of the spin--orbit tilt angles $\Delta i$ of the final binaries, assuming that the initial binary had spins aligned with the orbit. In the case of three-body encounters without the SMBH's tidal field, the tilt angle distribution is bimodal and favours both aligned and anti-aligned spins, which is consistent with the results of \citet{samsing2022}. However, once we introduce the tidal field, we begin to see deviations at high disc velocity dispersion.

Binary encounters in isolation are strongly bimodal, with two sharp peaks at $\Delta i \sim 0^\circ$ and $\Delta i \sim 180^\circ$. This is especially true at low velocity dispersions, which is easily explained by noting that $\sigma$ controls the inclination between binaries and singles. As $\sigma$ tends to zero, the system is increasingly constrained into a plane, meaning that the only two possible inclinations are aligned or anti-aligned.

On the other hand, in the presence of the SMBH, the tidal field can amplify the small differences in inclination between the binary and the single. Consequently, for $\sigma \geq 10^{-4}$, the sets with the SMBH have flatter distributions in $\cos (\Delta i)$, and the bimodality is entirely absent for $\sigma \geq 10^{-3}$.
As with the eccentricity distribution, the binaries from the isotropic set~\sset{-{$\cal MB$}} have a $\Delta i$ distribution that is closer to that of set~\sset{-1} than to that of set~\sset{-1-i}.

Therefore, our results show that we should only expect a completely bimodal distribution of spin--orbit misalignment in the most aligned and circular discs. Even a small disc velocity dispersion will prevent the formation of anti-aligned $\Delta i \sim 180^\circ$ binaries, which is because of the presence of the SMBH's tidal field.

\subsection{Caveats}
Here, we consider nuclear discs around a Milky-Way-like SMBH with different velocity dispersion. The velocity dispersion we chose is consistent with that arising from two-body interactions in a disc, and can be expressed as a single, dimensionless parameter $\sigma$ \citep{stewart2000}. However, the degree of velocity dispersion in a disc of BHs around an SMBH will depend on its formation mechanism. \REV{For example, a disc formed by the infall and disruption of a molecular cloud might be eccentric and lopsided following its formation \citep{map12,tra16,trani2018}. This configuration would likely lead to an excess of encounters at pericenter, where most of the orbits would cross, and at apocenter, where the bodies would spend most of their time. We plan to investigate this specific configuration in a future work.

For discs of black holes formed in an AGN disc, $\sigma$ is comparable to the disc-to-height ratio ($h/R$) of the  accretion disc. The typical $h/R$ for an accretion disc around a $10^{6}\rm\,\msun$ SMBH \citep[][see also D.Gangardt, in preparation]{sirko2003} is 0.1--$10^{-2}$, which corresponds to our \sset{1} and \sset{2} sets.

On the other hand, if the disc has formed through capture in an AGN gaseous disc, many BHs could have aligned and circular orbits ---depending on the thickness and lifetime of the disc--- because of gas drag forces \citep{nasim2022,generozov2023}. Our set~\sszero{} best represents the environmental conditions in a perfectly aligned and circular disc. Our results point out that encounters in discs with $\sigma \leq 10^{-5}$ produce outcomes that are approximately equivalent to those of a circular disc, because their velocity dispersion is dominated by the Keplerian shear only. A transition to a random-motion-dominated disc begins to emerge only at $\sigma \sim 10^{-4}$, which corresponds to eccentricities and inclinations of order $e,i \sim 0.000125, 0.0036^\circ$.

These small values of $e$ and $i$ can only be achieved in an AGN disc if the alignment and subsequent circularisation processes are exceptionally efficient. However, such efficiency is deemed unlikely due to the limited lifespan of AGNs, which allows only a fraction of orbiting bodies to align within the disc. Moreover, those bodies that do become aligned often retain substantial eccentricities \citep{generozov2023}. Additionally, the turbulence within the disc, which is responsible for driving the viscosity and mass accretion, introduces stochastic forces on embedded bodies, hindering complete circularisation. Depending on the density of the BH population, mutual gravitational interactions between BHs may prevail over gas drag, leading to an increase in velocity dispersion.
Based on these considerations, we argue that BH discs in AGNs should have large enough eccentricities and inclinations to make them random-motion dominated.
}

Additionally, if the disc was formed in situ, it will be composed of both stars and compact objects. Encounters between main sequence stars and BHs might lead to interesting phenomena, such as tidal disruption events, but we reserve the study of these for future work.

Throughout this work, we focused on comparing the outcome properties of three-body encounters, without considering the encounter rates between binaries and singles. While we find that three-body encounters in a disc result in more eccentric mergers than encounters in isolation, the frequency of these encounters remains to be determined. Migration traps in AGN discs have been invoked as a catalyst for such encounters \citep[e.g.][]{bellovary2016}, but the existence of such traps still needs to be fully ascertained \citep[e.g.][]{grishin}. In addition, the feasibility of aligning and embedding compact objects from around the SMBH within the lifetime of the AGN still needs to be explored using detailed AGN disc models (D.Gangardt, in preparation).

\REVV{Another crucial assumption in this work is that binaries lie in the same plane as their orbit around the SMBH. Recent hydrodynamical studies suggest that misaligned binaries embedded in AGNs may gradually align with the disc over time \citep{dittmann2024}, but the authors caution that this assumption may not hold for all types of discs around the SMBH.
}

Finally, the extent to which the dynamics of BHs in a dry disc apply to BHs embedded in an AGN disc remains unclear. We will explore this aspect in a follow-up work.

\section{Conclusion}
We investigated the outcomes of binary--single encounters in a disc of BHs around an SMBH by means of highly accurate simulations including up to 3.5 post-Newtonian corrections. We examined the effect of the disc velocity dispersion (Section~\ref{sec:vdisc}), the impact of the SMBH's tidal field (Section~\ref{sec:smbhtid}), and how these factors affect the properties of merging BH binaries (Section~\ref{sec:gw}). Our results can be summarised as follows:
\begin{enumerate}\itemsep5pt 
        \item The disc velocity dispersion $\sigma$, which parametrises the eccentricity and inclination of its constituent orbits, has a significant impact on the outcome of three-body encounters.
        For decreasing velocity dispersion, the eccentricity of post-encounter binaries goes from subthermal to superthermal, and binaries become harder (Figure~\ref{fig:aeenc}). This transition from thermal to superthermal eccentricity happens sharply at around $\sigma \approx 10^{-4}$. Encounters in discs with $\sigma \leq 10^{-5}$ yield outcomes that are almost the same as in perfectly circular and aligned discs, because for $\sigma \leq 10^{-5}$ the relative velocity of the binary--single pair is dominated by the Keplerian shear rather than by the velocity dispersion (Figure~\ref{fig:venc}).
        The main reason for this is that a higher velocity dispersion will enlarge the hard--soft boundary of the binaries embedded in the disc. Consequently, assessing the proper velocity dispersion of nuclear discs is key to correctly estimating the merger properties of their binary population.
        
        \item The impact of the SMBH's tidal field also depends on the disc velocity dispersion. By comparing with encounters in isolation, we find that neglecting the SMBH overestimates the number of mergers (by up to a factor of ${\sim}12$) for $\sigma \geq 10^{-4}$ and underestimates it (by up to a factor of ${\sim}2$) for $\sigma < 10^{-4}$ (Figure~\ref{fig:outcomp}). This occurs because, if we neglect the SMBH, the final binaries are harder and more eccentric relative to encounters in the presence of an SMBH, but only at high velocity dispersion (Figure~\ref{fig:smbh_iso}).
\end{enumerate}

Highly eccentric mergers are frequent during encounters in low-velocity-dispersion environments, when the binary and the single lie in the same plane, as proposed by \citet{samsing2022}. However, we find that this simple picture is complicated when considering the tidal field of the SMBH on encounters in discs with a velocity dispersion.

For high velocity dispersions $\sigma > 10^{-4}$, our results are in agreement with the outcomes of the disintegration of triples in a Keplerian potential \citep{trani2019b}, with these latter authors finding that the SMBH's tidal field hampers the production of a highly eccentric binary. A low disc velocity dispersion (i.e. dominated by the Keplerian shear only) reverses the role of the SMBH, and helps in the production of a superthermal distribution of final binaries.

The interplay between disc velocity dispersion and SMBH tidal field is also visible in the distributions of spin--orbit misalignment (Figure~\ref{fig:finincfig}). \REVV{For dispersion-dominated discs, ($\sigma > 10^{-4}$),} the SMBH's tidal field tends to isotropise the orientation of the final binaries, hampering the formation of binaries with anti-aligned spins.

Regardless of the disc velocity dispersion, we show that a nuclear disc can still produce a larger number of mergers than NCSs that do not host an SMBH, \REVV{as long as the binaries lie in the plane of the disc. Even the disc with  the highest velocity dispersion considered here ($\sigma=10^{-1}$, $e,i \sim 0.125, 3.6^\circ$, equivalent to $\sigma_{\cal MB} \sim 50 \,\rm km/s$) has approximately twice the number of mergers compared to an NSC without a SMBH, where the encounters show no preference in regard to the orientation of the binaries with respect to the incoming singles}. 

Our work points out the importance of characterising the velocity dispersions of nuclear BH discs ---whether they are formed via AGN interaction or via other mechanisms--- in order to constrain the properties of GW sources in galactic nuclei.
In our next work, we will perform a demographic study of merging BH binaries in a nuclear disc, taking into account binary--single encounter rates and the interplay with ZKL oscillations, \REV{even after the encounters have taken place}.

\begin{acknowledgements}
   We would like to thank the anonymous referee for multiple useful comments and suggestions that helped to improve this work. A.A.T. wishes to thank Evgeni Grishin and Rosemary Mardling for helfpul comments and criticism.
   A.A.T. acknowledges support from JSPS KAKENHI Grant Number 21K13914 and from the European Union’s Horizon 2020 and Horizon Europe research and innovation programs under the Marie Sk\l{}odowska-Curie grant agreements no. 847523 and 101103134. All the simulations were performed on the \texttt{awamori} computer cluster at The University of Tokyo.
\end{acknowledgements}

\bibliographystyle{aa}
\bibliography{bhdisk,totalms}

\end{document}